\documentclass[aps,prb,amsmath,amssymb,twocolumn,superscriptaddress]{revtex4-1}

\usepackage{graphicx}
\usepackage{dcolumn}
\usepackage{bm}
\usepackage{hyperref}

\begin{document}

\title{Doping and temperature evolution of pseudogap and spin-spin correlations in the two-dimensional Hubbard model}

\author{V. I. Kuz'min}
\email{kuz@iph.krasn.ru}
\affiliation{Kirensky Institute of Physics, Federal Research Center KSC SB RAS, Krasnoyarsk, 660036 Russia}

\author{M. A. Visotin}
\affiliation{Kirensky Institute of Physics, Federal Research Center KSC SB RAS, Krasnoyarsk, 660036 Russia}

\author{S. V. Nikolaev}
\affiliation{Kirensky Institute of Physics, Federal Research Center KSC SB RAS, Krasnoyarsk, 660036 Russia}
\affiliation{Siberian Federal University, Krasnoyarsk, 660041 Russia}

\author{S. G. Ovchinnikov}
\affiliation{Kirensky Institute of Physics, Federal Research Center KSC SB RAS, Krasnoyarsk, 660036 Russia}
\affiliation{Siberian Federal University, Krasnoyarsk, 660041 Russia}

\date{\today}

\begin{abstract}
Cluster perturbation theory is applied to the two-dimensional Hubbard $t-t'-t''-U$ model to obtain doping and temperature dependent electronic spectral function with $4 \times 4$ and 12-site clusters. It is shown that evolution of the pseudogap and electronic dispersion with doping and temperature is similar and in both cases it is significantly influenced by spin-spin short-range correlations. When short-range magnetic order is weakened by doping or temperature and Hubbard-I like electronic dispersion becomes more pronounced, the Fermi arc turns into large Fermi surface and the pseudogap closes. It is demonstrated how static spin correlations impact the overall dispersion's shape and how accounting for dynamic contributions leads to momentum-dependent spectral weight at the Fermi surface and broadening effects.

\end{abstract}

\maketitle

\section{\label{Intro}Introduction} 
Revealing the nature of high-$T_c$ superconductivity in cuprates is one of the major challenges in condensed matter physics. One step towards solution of the high-$T_c$ problem is to understand the behavior of the normal state electronic structure. Thus, the pseudogap \cite{Timusk} phase located below the temperature $T^*\left(p\right)$ decreasing with doping $p$ and its relation with high-$T_c$ superconductivity have gained a lot of attention. A pseudogap metal is considerably different from weakly correlated metals described by weak-coupling perturbation theory \cite{AGD}. The primary nature of the pseudogap, if there exists one, is a highly debated topic with many candidates \cite{Norman05, Taillefer10, Keimer, Kordyuk, Vishik18, Fratino16}.  

The 2D Hubbard model \cite{Hubbard} is believed to possess the main ingredients of the cuprate layers’ low-energy properties. Due to growing precision of experimental data like electronic spectra obtained by angle-resolved photoemission spectroscopy (ARPES) \cite{Damascelli, Kordyuk, Yoshida12, Vishik18}, measurements of quantum oscillations in cuprates \cite{DL07, Barisic13, DL15}, and other techniques \cite{Renner98, McElroy05, Zheng, Kawasaki, LeBoeuf07, LeBoeuf11, DL13}, new peculiarities of the pseudogap come to knowledge over time. This way, the electronic structure and the pseudogap behavior in the Hubbard model and its low-energy $t-J$ models \cite{Chao} have been revisited many times and studied by a number of different numerical approaches such as quantum Monte-Carlo (QMC)\cite{Bulut, Preuss, Preuss97, Grober, Moritz09}, cluster perturbation theory (CPT) \cite{Senechal00, Senechal02}, variational cluster approximation (VCA) \cite{Potthoff03}, dynamical mean-field theory (DMFT) \cite{Georges} and its cluster (CDMFT) \cite{Hettler, Maier05} and diagram \cite{Sadovskii, Rohringer18} extensions, and other techniques designed for dealing with strongly correlated systems \cite{Zaitsev, Plakida07, Avella07, Korshunov, Sherman18}. A large body of theoretical work is concentrated on the doping evolution of the pseudogap, while its temperature dependance still lacks a systematic investigation.

From the studies of the pseudogap within the 2D Hubbard and the $t-J$ models we know that short-range antiferromagnetic (AFM) correlations, which are their distinctive properties, should be the main origin of the pseudogap within these models. It is of interest to make a qualitative comparison of the temperature evolution of the electronic spectral function in the 2D Hubbard model with the main trends in recent experimental data obtained by ARPES in order to clarify the role of electronic correlations in the physics of the pseudogap observed in real compounds. ARPES results show that a clear nodal-antinodal distinction exists at low temperature and doping, and that the Fermi surface is arc-like due to an influence of the dramatic change of electronic self-energy from nodal to antinodal directions \cite{Li18}. A growth of the arc with temperature has been reported \cite{Norman98, Kanigel, Reber12}, which is similar to its well-known growth with doping. Recent ARPES results lead to an intriguing conclusion that at least one critical temperature exists above $T_c$ within the pseudogap phase \cite{Kordyuk, Vishik18}. It is fascinating to investigate whether the electronic correlation physics of the Hubbard model can be relevant to this non-monotonous behavior.

In this paper we study doping and temperature evolution of the electronic spectrum in the 2D Hubbard model using CPT. The spectral function is examined along with intracluster static spin-spin correlation functions, which influence the spectrum significantly, and thus provide information about the relationship of short-range magnetic order with the electronic spectral properties. First, in agreement with several previous studies we observe that the evolution of the low-energy electronic structure from low to high doping proceeds through three regions. Within the pseudogap state we identify a strong pseudogap (SPG) state at very low doping and a weak pseudogap (WPG) state at higher doping. These terms have been previously used in the literature \cite{Schmalian} with different meaning but seem appropriate in our case. At larger doping compared to the WPG the pseudogap is closed and the state similar to a normal Fermi liquid (NFL) is observed. Second, we show that the temperature evolution of the electronic structure and the pseudogap in the 2D Hubbard model has a great similarity to its evolution with doping. When spin-spin correlations are weakened by temperature, a Hubbard-I like dispersion develops in agreement with QMC results \cite{Preuss97, Grober} (see also Ref.~\onlinecite{Wang18} for the dependence of this feature on doping and superexchange within CPT) similarly as it does with doping. As a consequence, the nodal-antinodal distinction diminishes around the Fermi level and the Fermi arc turns into a large Fermi surface. We find that in this case the evolution of the electronic stucture goes through the same three stages.

Since CPT has been applied mainly for the case of zero temperature, we hope that our result will provide a new and useful reference for future studies at finite temperatures. As it will be presented below, our temperature-dependent spectra reveal some qualitative similarities with temperature-dependent ARPES spectra, pointing again at the important role played by short-range AFM in the physics of the pseudogap. We bring attention that here no attempt is made to draw a phase diagram of cuprates since the electron-phonon interaction is not taken into account. Thus we do not obtain charge density waves (CDW), which have been realized to be an important part of physics of cuprates \cite{Fradkin}.

The rest of this paper is organized as follows. In Sec.~\ref{sec:2} we briefly discuss the method. Section~\ref{sec:3} is devoted to the presentation of results. In Sec.~\ref{sec:4} we discuss the results obtained. In Sec.~\ref{sec:5} concluding remarks are given. The details about the calculations can be found in Appendixes~\ref{sec:a1}, \ref{sec:a2}, and \ref{sec:a3}.

\section{\label{sec:2} Model and method}

The Hubbard model \cite{Hubbard} is given by the Hamiltonian  
\begin {equation}
H=\sum\limits_{i,\sigma }{\left\{{\left({\varepsilon-\mu} \right){n_{i,\sigma }}+\frac{U}{2}{n_{i,\sigma }}{n_{i,\bar\sigma}}} \right\}-\sum\limits_{ij,\sigma} t_{ij} a_{i,\sigma }^{\dag} a_{j,\sigma}^{}},
\label {eq:1}
\end {equation}
where $a_{i,\sigma }$ denotes the annihilation operator of an electron on a site $i$ with spin $\sigma$, the particle number operator is $n_{i\sigma}=a_{i\sigma}^{\dag}a_{i\sigma}^{}$, $t_{ij}$ is the hopping integral, and $U$ is the on-site Coulomb interaction.

One of the approaches used to study the electronic structure of the Hubbard model is CPT. The idea behind it is to incorporate long-range interactions by means of perturbation theory into the data obtained exactly within an isolated cluster. The CPT approximation can be obtained by accounting for the first order hopping process within the strong coupling perturbation theory \cite{Pairault, Senechal00, Senechal02} or shown to be the generalized Hubbard-I approximation within the X-operator perturbation theory \cite{Nikolaev10, Nikolaev12} with Hubbard X-operators constructed on the basis of the exact eigenstates of a finite cluster including all intracluster correlations. Within CPT, ARPES like spectra are obtained with continuous momentum resolution. Another virtue of the method is that it enables treating larger clusters than more sophisticated cluster approaches when using exact diagonalization (ED) technique, which gives access to the Green functions defined in real frequency space, and provides calculations’ complexity independence on non-nearest hopping parameters and doping.  Whereas CPT has been extensively used to study doping dependent electronic structure of models of strong electronic correlations at zero temperature \cite{Senechal00, Senechal02, Senechal04, Kohno12, Kohno14, Kohno15, Yang, Wang15, Wang18, Ivantsov17, Ivantsov18} and applied several times at finite temperatures \cite{Aichhorn03, Kawasugi16}, to our knowledge there exists no detailed investigation of the temperature dependence of the pseudogap within CPT. Although CPT is not a self-consistent method (contrary to VCA or CDMFT) and thus cannot be used by itself to study ordered phases, since the pseudogap is a normal state phenomenon, CPT is fully applicable in our case.

In this paper we study the doping evolution of the electronic spectral function using a $4\times4$ cluster at zero temperature. We confine ourselves to the doping range $0.03 \leq p \leq 0.25$ due to significant influence of finite-size effects at large doping levels from the one side and since all our calculations are carried out for the paramagnetic state not capable of describing a very low doping state adequately enough from the other side. The temperature dependence is studied using a 12-site cluster that preserves point symmetry of a square. The details concerning our implementation of CPT are given in Appendix~\ref{sec:a1}.

\section{\label{sec:3} Results}

\subsection{Case of zero doping and non-nearest neighbor hoppings}
\begin{figure}
\includegraphics{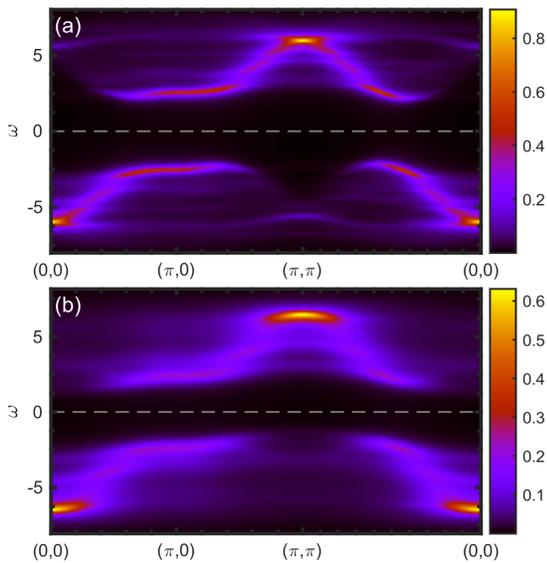}
\caption{\label{fig:1} The electronic spectral function at half filling at (a) $\beta = 10/t$, (b) $\beta = 3/t$ obtained with a 12-site cluster. The chemical potential is at zero energy here and below.}
\end{figure}

First of all we consider the case $p=0$, $t'=0$, and $t''=0$, since it gives us an opportunity to compare the results with quantum Monte Carlo data, available for these parameters at low temperatures \cite{Grober} and estimate the adequacy of our calculations. The temperature dependence of the spectrum is seen by comparing Fig.~\ref{fig:1} (a) and (b) for the inverse temperature $\beta = 10/t$ and $\beta = 3/t$. A characteristic feature is that at $\beta = 10/t$, in the lower Hubbard band, the major spectral weight residing at point $\left(\pi, \pi\right)$ is concentrated at high energy $\omega \approx -6t$, and significant amount of the low-energy $\omega \approx -2t$ spectral weight is concentrated near $\left(\pi/2, \pi/2\right)$, thus the spectrum contains the influence of strong AFM fluctuations. The redistribution of $\left(\pi, \pi\right)$-point spectral weight to lower energies with heating is seen at $\beta = 3/t$ meaning that the Hubbard-I like dispersion becomes pronounced in accordance with Ref.~\onlinecite{Grober}. Already at quite low temperature $\beta = 10/t$ the short-range AFM gives qualitatively similar picture to the QMC results where quasi long-range order is present, at $\beta=3/t$ the qualitative agreement between these two methods becomes even better, meaning that the CPT results improve with increasing temperature. The spectral map at $\beta=3/t$ is very similar to the result obtained by the variational approximation in the paramagnetic phase \cite{Seki, Nishida20} at the same temperature.

\subsection{Case of finite doping and non-nearest neighbor hoppings}
Here we consider the case of more realistic parameters for cuprates when the influence of the second and the third neighbors is included. Our main set of hopping parameters will be $t' = -0.2t$, $t'' = 0.15t$.

\begin{figure}
\includegraphics{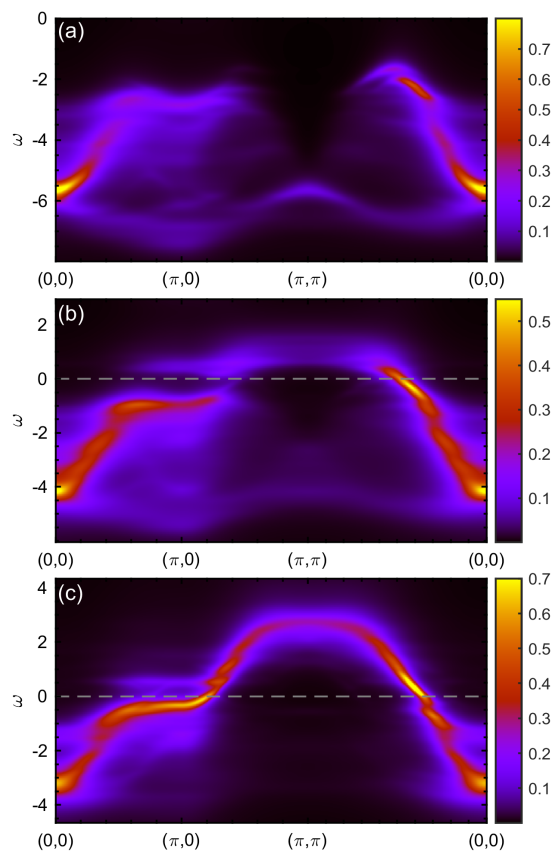}
\caption{\label{fig:2} The electronic spectral function at $T=0$ obtained using a $4\times4$ cluster for different values of doping: (a) $p=0$, (b) $p=0.0625$, (c) $p=0.25$.}
\end{figure}

\begin{figure}
\includegraphics{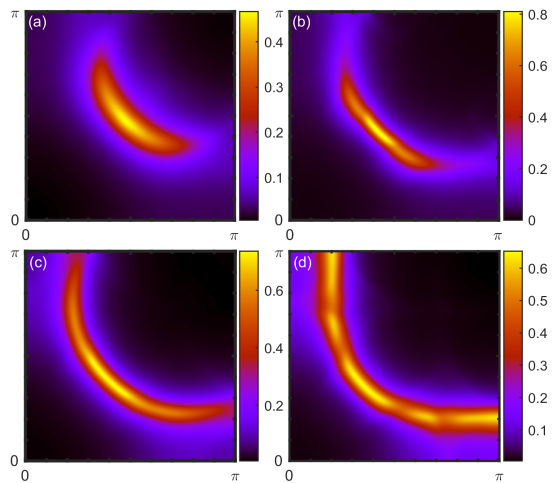}
\caption{\label{fig:3} The electronic spectral function map at the Fermi level at $T=0$ obtained using a $4\times4$ cluster. The value of doping is (a) $p=0.0625$,  (b) $p=0.125$, (c) $p=0.1875$, (d) $p=0.25$.}
\end{figure}

Figure~\ref{fig:2} shows the dispersion of the lower Hubbard band of electrons that corresponds to the valence band of a hole doped cuprate. One can see that a feature similar to the Hubbard-I dispersion emerges with doping. A pronounced signature of this behavior is that the spectral weight at $(\pi,\pi)$ disappears below the Fermi level, and the dispersion above it gains coherence in accordance with Ref.~\onlinecite{Wang18}. As a consequence the Fermi surface gradually turns from the small Fermi arc in Fig.~\ref{fig:3}(a) into the large full Fermi surface in Fig.~\ref{fig:3}(d). The pseudogap is seen as a dip in spectral weight around the Fermi level in the $(\pi,0)-(\pi,\pi)$ direction in Fig.~\ref{fig:2}(b), while at large doping in Fig.~\ref{fig:2}(c) such dip is absent. Note that there are further dips of the spectral weight, clearly seen in Fig.~\ref{fig:3}(d) along the Fermi surface. These are not associated with the pseudogap but result from artificial density wave formation due to scattering by the reciprocal vectors of the cluster superlattice, which is inherent in cluster methods (see Ref.~\onlinecite{Verret19} for more details).

\begin{figure}
\includegraphics{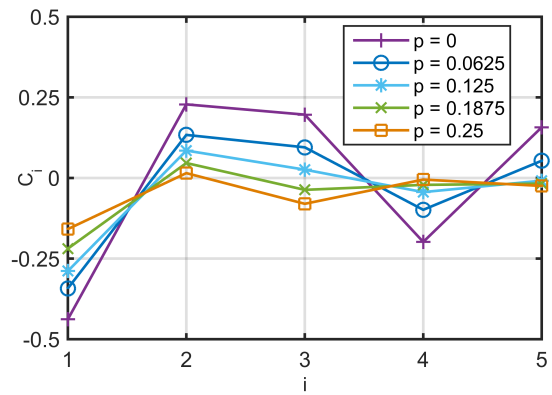}
\caption{\label{fig:4} Spin-spin correlation function defined in Eq.~\ref{eq:2} for different values of doping at $T=0$.}
\end{figure}

\begin{figure}
\includegraphics{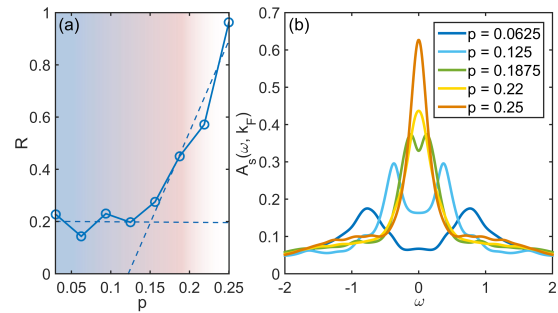}
\caption{\label{fig:5} (a) The antinodal/nodal ratio $R$ for the spectral weight at the Fermi level, defined in the text for different doping levels; the two dashed lines are linear fits to data within the SPG doping range and outside; smooth transitions between background colors illustrate different doping regimes discussed in the text} (b) the symmetrized spectral function defined by Eq.~\ref{eq:3} for a wave vector $\textbf{k}_F$ in the antinodal direction for different values of doping at zero temperature (finite temperature $\beta = 12/t$ was substituted into the Fermi-Dirac function in Eq.~\ref{eq:3} to produce smooth ARPES-like curves)
\end{figure}

In Fig.~\ref{fig:4} we show the doping evolution of intracluster spin-spin correlation functions
\begin {equation}
C_i = \left\langle \left(n_{\uparrow a}-n_{\downarrow a}\right)\left(n_{\uparrow b}-n_{\downarrow b}\right)\right\rangle_{\left|\mathbf{R}_a-\mathbf{R}_b\right|\in i},
\label {eq:2}
\end {equation}
where $\left\langle...\right\rangle_{\left|\mathbf{R}_a-\mathbf{R}_b\right|\in i}$ denotes that the correlation functions are additionally averaged among sites $a$ and $b$, the distance between which $\left|\mathbf{R}_a-\mathbf{R}_b\right|$
belongs to the $i$-th coordination sphere. We stress that the correlation functions in this work are calculated not within CPT but within a cluster with open boundary conditions (as in CPT), thus their influence is contained in the CPT spectra. The dispersion in Fig.~\ref{fig:2} and the doping evolution of the Fermi contours in Fig.~\ref{fig:3} should be considered together with the doping dependence of the spin correlation functions. Analysis of spin correlators $C_i$ in Fig.~\ref{fig:4} as a function of distance in terms of coordination sphere number $i$ for different doping levels up to $p=0.25$ indicates that AFM short-range order that has $C_1<0$, $C_2$ and $C_3>0$, $C_4<0$, $C_5>0$ takes place for doping $p=0; 0.0625; 0.125$. At zero doping a strong short-range AFM ordering with tendency to long range, which is violated by a finite size of a cluster, is observed. For small doping the influence of AFM is clearly pronounced in the dispersion shape presented in Fig.~\ref{fig:2}(a), (b). If one considers the antinodal/nodal spectral weight ratio at the Fermi level $R = A_{AN}\left(\mathbf{k}_F,\omega=0\right) / A_{N}\left(\mathbf{k}_F,\omega=0\right)$, one can conclude from Figs.~\ref{fig:3} and \ref{fig:5}(a) that for $p \leq 0.125$ the value of $R$ is rather small and practically constant. Possibly, non-zero value of this ratio is a consequence of CPT artifacts. For convenience of the following discussion we call this state at low doping ``the strong pseudogap'' (SPG). In ARPES it is quite common to study the change of the pseudogap using the symmetrized spectral function $A_{s}\left(\omega,\textbf{k}_F\right)$:
\begin {equation}
A_{s}\left(\omega,\textbf{k}_F\right) = f\left(\omega,\beta\right)A\left(\omega,\textbf{k}_F\right) + f\left(-\omega,\beta\right)A\left(-\omega,\textbf{k}_F\right),
\label {eq:3}
\end {equation}
where $\textbf{k}_F$ is defined as a point of the maximal spectral weight in the antinode and $f\left(\omega,\beta\right)$ is the Fermi-Dirac distribution function. Indeed, the antinodal spectra in Fig.~\ref{fig:5}(b) shows a pronounced pseudogap behavior at low doping, so SPG is an adequate term in this case.

For large doping $p = 0.1875$ the short-range order is already changed to $C_1<0, C_2>0, C_3<0$ and more distant practically zero: the third correlations change sign, which may be a manifestation of the Nagaoka ferromagnetism \cite{Nagaoka}. This type of correlations violates the short-range AFM order. At the same time a noticeable growth of $R$ is seen starting from $p = 0.125$ in Fig.~\ref{fig:5}(a). However, Fig.~\ref{fig:5}(b) shows that for doping $p \lesssim 0.2$ the pseudogap state is realized. Note that already for $p=0.1875$ the Fermi surface seems to be close to Fermi-liquid, damping in the antinodal direction is not very strong. Thus, the spectral weight distribution and short-range order are qualitatively different from SPG. We will use the term ``weak pseudogap'' (WPG) for this doping region. In our calculations the crossover between SPG and WPG appears as a smooth transition between $p = 0.125$ and $p = 0.1875$. Finally, at $p \gtrsim 0.2$ the pseudogap is closed (see Fig.~\ref{fig:5}(b)), for $p=0.25$ a remnant of short-range AFM is seen only for the first coordination sphere in Fig.~\ref{fig:4}, and the large Fermi surface is observed in Fig.~\ref{fig:3}(d). Such behavior is typical for a Fermi liquid, so we will call the doping region at $p \gtrsim 0.2$ the normal Fermi liquid (NFL). Similar evolution of the Fermi arcs has been obtained earlier within cluster DMFT \cite{Civelli05, Stanescu06, Haule07, Sakai09, Ferrero, Ferrero09}, composite operators approach \cite{Avella14}, it is in general agreement with the doping dependence of the electronic structure within dynamical cluster approximation (DCA) \cite{Gull08, Gull09, Gull10}; the growth of Fermi arc with doping is well-known from ARPES data \cite{Shen05, Yoshida06, Yoshida09}.

\begin{figure}
\includegraphics{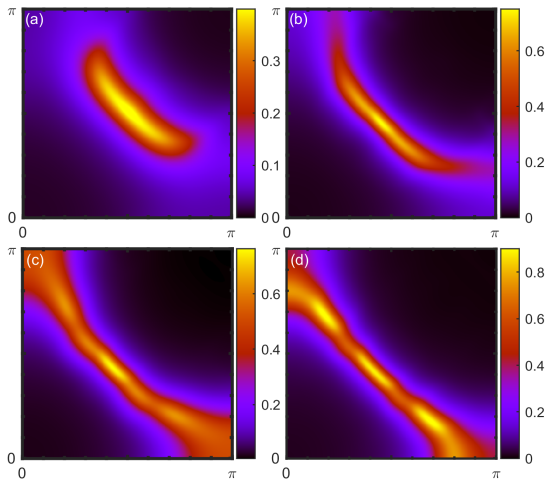}
\caption{\label{fig:6} The same as in Fig.~\ref{fig:3} for $t''=0$.}
\end{figure}

We also investigate the doping evolution of the electronic structure for a different set of parameters by setting $t''=0$, since third-neighbor hopping processes influence the spectrum significantly by stabilizing the dispersion in the nodal direction. The case described above ($t''=0.15$) is qualitatively reminiscent of the electronic structure of such compounds as $\text{Tl}_2\text{Ba}_2\text{CuO}_{6+\delta}$ and $\text{YBa}_2\text{Cu}_3\text{O}_{7-\delta}$, where no transition from a hole-like Fermi surface around $(\pi,\pi)$ to an electron-like around $(0,0)$ is observed even at large doping \cite{Plate05, Peets07, Hossain08}, as opposed to the case of zero third-neighbor hoppings: in Fig.~\ref{fig:6} at $p=0.0625$ and $p=0.125$ the zero-frequency spectral weight is very similar to the previously considered, but at larger doping differs significantly. At $p=0.25$ the spectral weight constitutes a feature very similar to the electron-like pocket observed in $\text{La}_{2-x}\text{Sr}_x\text{CuO}_4$ \cite{Yoshida01, Yoshida06, Razzoli10}.

\begin{figure}
\includegraphics{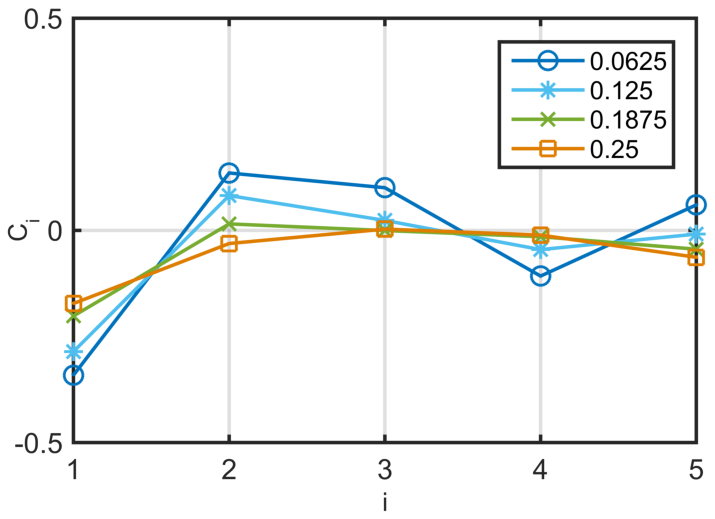}
\caption{\label{fig:7} The same as in Fig.~\ref{fig:4} for $t''=0$.}
\end{figure}

\begin{figure}
\includegraphics{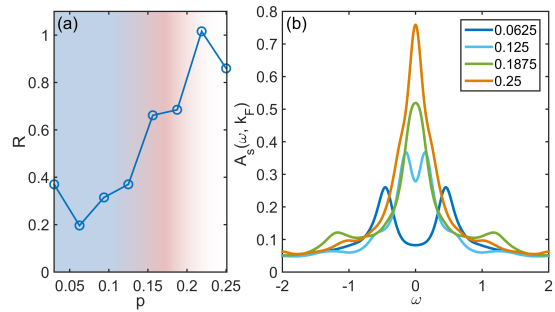}
\caption{\label{fig:8} The same as in Fig.~\ref{fig:5} for $t''=0$.}
\end{figure}

Spin correlations at $t''=0$ are almost the same as for the previously used parameters up to $p=0.125$. At larger doping the short-range AFM within a cluster is destroyed very quickly (see Fig.~\ref{fig:7}). Similarly to the case of $t''=0.15$, the onset of rapid growth of antinodal/nodal spectral weight ratio in Fig.~\ref{fig:8}(a) coincides approximately with the doping region where the short-range AFM fades. We conclude that at $p<0.125$ the SPG is observed: the Fermi level is in the vicinity of the bottom of the pseudogap, and minimum in Fig.~\ref{fig:8}(a) is approximately its lower point. The pseudogap now closes at much lower doping (see Fig.~\ref{fig:8}(b)), so the WPG is in the narrow range $0.125 \lesssim p\lesssim0.15$. In general the evolution of the electronic structure in this case is in agreement with the previously considered. The AFM short-range order is destroyed more quickly outside the SPG range: already for $p=0.1875$ the second neighbor spins are not aligned antiferromagnetically. The pseudogap closes at smaller doping as well.

\begin{figure}
\includegraphics{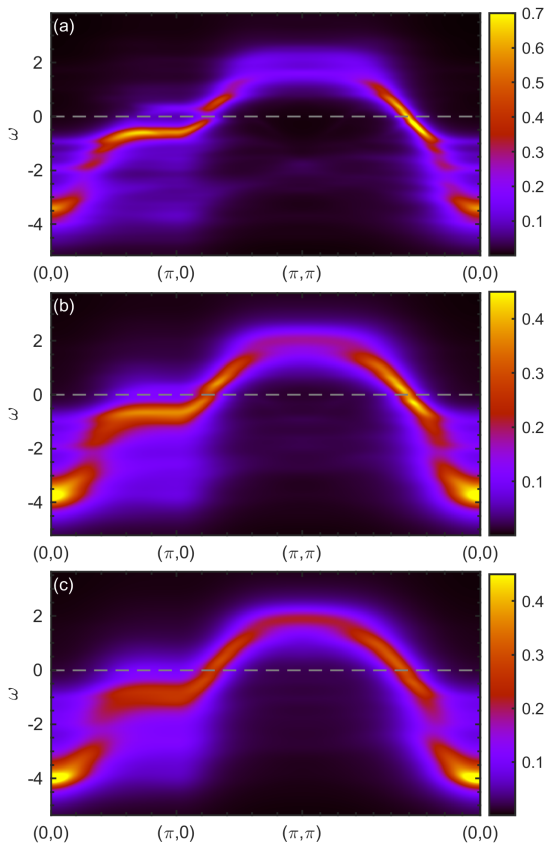}
\caption{\label{fig:9} The electronic spectral function at $p=0.167$ obtained using a 12-site cluster at (a) $\beta=24/t$, (b) $\beta=8/t$, (c) $\beta=4/t$. }
\end{figure}

Let us turn to the investigation of temperature evolution of spectral function studied with a 12-site cluster at doping $p=0.167$ (more precisely, $p=1/6$ with 10 electrons per cluster). For the following we use the main set of parameters $t'=-0.2$, $t''=0.15$. In Appendix~\ref{sec:a2} we present a comparison of the results obtained at this value of doping and zero temperature with 16-site and 12-site clusters to show that a 12-site cluster does not introduce major discrepancies with respect to the results obtained with a $4\times 4$ cluster and that SPG is still observed for this value of doping for a 12-site cluster. The main trends of transformation of the spectral function with increasing temperature are seen in Fig.~\ref{fig:9}. The region around point $(\pi,\pi)$ above the Fermi level becomes more coherent and the feature similar to a bare dispersion becomes more pronounced. However, the waterfall-like feature at high energies is stable at high temperatures. The pseudogap gradually closes, as seen from the antinodal cut: it is well pronounced at $\beta=24/t$ and its signature is still visible at $\beta=8/t$, while it is absent at $\beta =4/t$.

\begin{figure}
\includegraphics{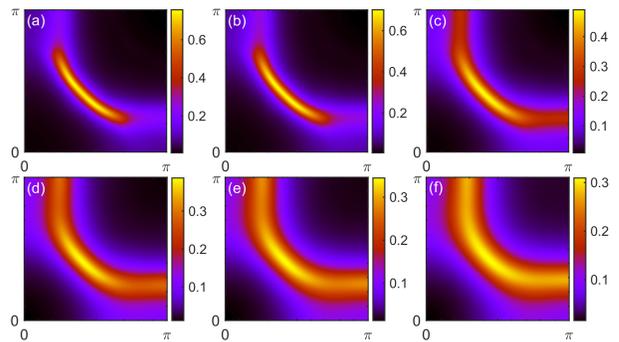}
\caption{\label{fig:10} The electronic spectral function map at the Fermi level at $p=0.167$ obtained using a 12-site cluster at (a) $\beta=10^{4}/t$, (b) $\beta=24/t$, (c) $\beta=12/t$, (d) $\beta=8/t$, (e) $\beta=6/t$, (f) $\beta=4/t$.}
\end{figure}

\begin{figure}
\includegraphics{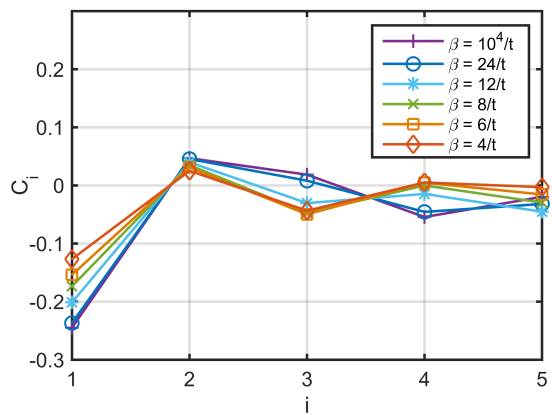}
\caption{\label{fig:11} Spin-spin correlation function defined in Eq.~\ref{eq:2} for different values of temperature at $p=0.167$.}
\end{figure}

\begin{figure}
\includegraphics{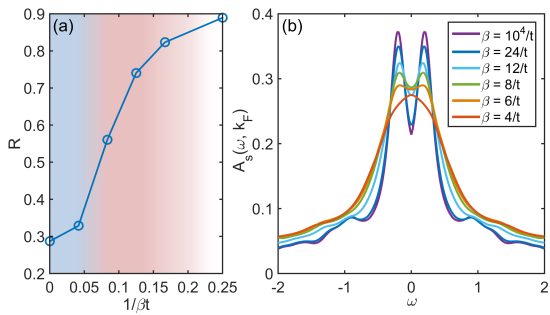}
\caption{\label{fig:12} (a) The antinodal/nodal ratio $R$ for the spectral weight at the Fermi level, defined in the text for different temperatures, (b) Symmetrized spectral function defined by Eq.~\ref{eq:3} for a wave vector $\textbf{k}_F$ in the antinodal direction at different values of temperature. Both are shown for $p=0.167$.}
\end{figure}

Let us consider the electronic structure evolution with heating in Figs.~\ref{fig:9},~\ref{fig:10} together with spin correlators in Fig.~\ref{fig:11}. In Fig.~\ref{fig:10} the growth of the Fermi arc with temperature is seen in qualitative agreement with ARPES \cite{Norman98, Kanigel, Reber12}. Similar reconstruction of Fermi arcs has been also obtained within the large-$N$ mean-field theory of the $t-J$ model due to short-range $d$-CDW fluctuations \cite{Bejas11}. With heating spin correlations are decreasing similarly to the case of doping. For $\beta = 10000/t$ and $\beta = 24/t$ the similar type of AFM short-range order with $C_1<0$, $C_2$ and $C_3>0$, $C_4<0$ and $C_5>0$ takes place. The antinodal/nodal spectral ratio in Fig.~\ref{fig:12}(a) is quite close at both temperatures. Thus, the SPG state is observed. At $\beta = 12/t$ the AFM correlations are seen only for the first and the second neighbors. For the third they change sign and become practically zero further. The spectral function shows WPG behavior with small finite $A_{AN}\left(\mathbf{k}_F,\omega=0\right) \ll A_{N}\left(\mathbf{k}_F,\omega=0\right)$, similar to $\beta=8/t$. We conclude that transformation from SPG to WPG goes between $\beta = 24/t$ and $\beta = 12/t$, as it is also seen from the growth of the antinodal/nodal spectral ratio in this temperature region. A transformation to the NFL seems to occur close to $\beta=6/t$. These conclusions are in agreement with the temperature dependence of the symmetrized antinodal spectral function in Fig.\ref{fig:12}, discussed below. Note also that due to a very similar behavior of spin correlations with doping and temperature the overall dispersion shape, which is affected a lot by static spin correlations, at $p=0.167$ and $\beta=4/t$ (Fig.~\ref{fig:9}(c)) is very similar to the obtained at $p=0.25$ and zero temperature (Fig.~\ref{fig:2}(c)). However, at $p=0.167$ and $\beta=4/t$ the spectrum is significantly broader due to different behavior of dynamical contribution to the correlation functions, which introduces effects of finite quasiparticle lifetime (see Ref.~\onlinecite{Plakida07}).

Considering the change of the symmetrized spectral function with temperature, for $\beta = 4/t$ a single peak typical for NFL is seen. The dip at $\omega = 0$ almost disappears at $\beta = 6/t$, which is a sign of the pseudogap formation temperature $T^*$\cite{Norman98, Kanigel}. For $\beta = 8/t$ and $\beta = 12/t$ the pseudogap deep at the Fermi level is clear but small, the value $A_{s}\left(\omega=0,\textbf{k}_F\right)$ is still smaller then the NFL maximum, so the term WPG seems to be appropriate. With further temperature decrease (for $\beta = 24/t$ and $\beta = 10000/t$) the PG deep is sharp and for these temperatures the SPG may be considered.

\section{\label{sec:4} Discussion}
\begin{figure*}
\includegraphics{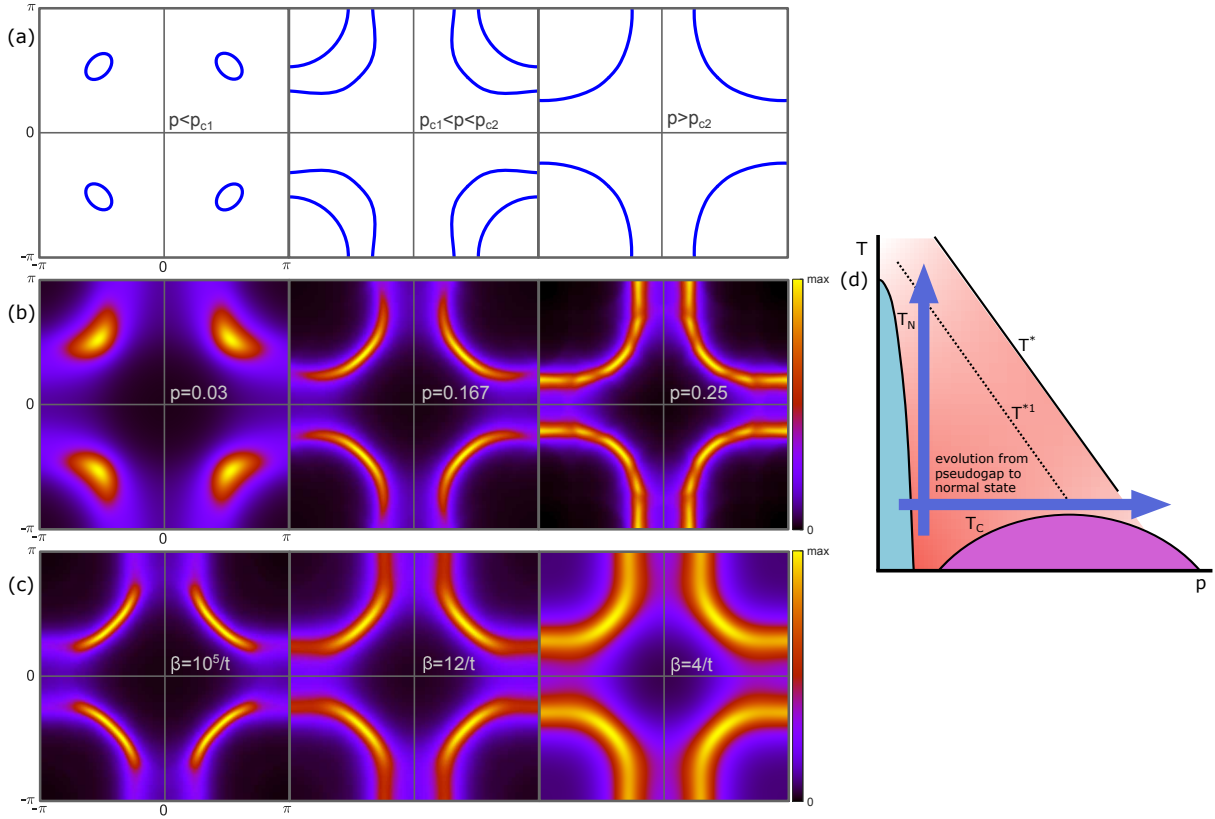}
\caption{\label{fig:13} (a) The schematic representation of the Fermi surface within static approximation for three values of doping divided by two Liphsitz transitions, (b) the CPT result using a $4\times4$ cluster at $T=0$ for similar doping values, (c) the CPT result using a 12-site cluster at $p=0.167$ for different temperatures, (d) sketch of the phase diagram, where arrows denote two directions of evolution of the electronic structure studied in this paper, $T_N$ is the AFM Neel temperature, $T_C$ is the superconducting transition temperature, $T^*$ is the pseudogap transition line (WPG to NFL transition), and $T^{*1}$ is the SPG to WPG crossover line.}
\end{figure*}

Summarizing our results we compare the doping and temperature evolution of the Fermi arcs in Fig.~\ref{fig:13}(b) and (c) with the old picture of the Lifshitz transitions with doping obtained within the generalized mean field for strongly correlated systems \cite{Korshunov, Ovchinnikov09, Ovchinnikov11}. Within the static approximation for the spin correlation function the imaginary part of the electronic self energy within the $t-J$ model is absent, while the real part results in the doping dependent electronic structure. The doping evolution of the Fermi surface in static approximation is schematically given in Fig.~\ref{fig:13}(a): Three doping regions have been obtained. At low doping the Fermi surface is given by 4 small Fermi surface pockets centered near $\left(\pi/2,\pi/2\right)$, these pockets increase their volume and touch the Brillouin zone boundary at some critical doping value $p_{c1}$ ($p_{c1}=0.16$ for the parameters chosen in that study), where the Lifshitz transition with the topology change of the Fermi surface occurs. Above $p_{c1}$ two Fermi surface pockets are centered around $\left(\pi,\pi\right)$ and the inner small pocket disappears at the second Lifshitz transition point $p_{c2}=0.24$. Above $p_{c2}$ there is one large Fermi surface around $\left(\pi,\pi\right)$ that corresponds to the NFL \cite{Ovchinnikov11}. In this approximation all points along the Fermi contour have equal spectral weight. Similar approach within the Hubbard model provides some non uniform distribution of the spectral weight along the Fermi contour due to the quasiparticle scattering between two Hubbard subbands \cite{Makarov19}.

Nevertheless, this non uniform spectral weight distribution does not change the closed pocket to the Fermi arc. This picture contradicts to the ARPES data where only arcs have been found with different lengths dependent on doping and temperature. This transformation occurs only when the electronic self-energy removes the spectral weight at the large part of a Fermi surface contour, as it has been shown within CDMFT and CPT \cite{Stanescu06, Sakai09, Sakai10, Nikolaev12} in terms of poles and zeros of the Green function. Similarly, in our calculations the Fermi arc growing with doping and temperature is observed. We consider the doping dependence in Fig.~\ref{fig:13}(b) to show also three concentration regions, the SPG, the WPG and the NFL with two transitions between them. The borders between these regions correspond to the critical concentrations of the Lifshitz transitions. We note that the conclusion about SPG-WPG and WPG-NFL transitions with doping is in qualitative agreement with the DCA studies of the electronic structure in the 2D Hubbard model done for the value of interaction similar to ours \cite{Gull09, Gull10}: the WPG-SPG transition should roughly correspond to the transition between the momentum space differentiated region and the sector selective phase (in terms of Ref.~\onlinecite{Gull10}), whereas WPG-NFL corresponds to the transition when the system enters the isotropic Fermi-liquid regime. The same type of the Fermi surface evolution with heating is shown in Fig.~\ref{fig:13}(c). Probably a transformation of the SPG to the WPG at $T^{*1}$ and from WPG into NFL at $T^*$, schematically depicted in Fig.~\ref{fig:13}(d), are smooth crossovers due to dynamical damping of the quasiparticles at the Fermi level.

\section{\label{sec:5} Conclusion}
In conclusion, we have studied the doping and temperature evolution of the electronic spectral function in the 2D Hubbard model on a square lattice using CPT, focusing on the pseudogap. The doping evolution has been studied at fixed temperature $T=0$, the temperature evolution has been investigated at fixed doping $p=0.167$. Together with the spectral function we have considered the doping and temperature dependence of the spin correlation function as a function of the intersite distance. In support of the previous studies done within the static approximation for the self-energy \cite{Korshunov, Ovchinnikov09, Ovchinnikov11} we see that short-range spin correlations provide a decisive influence on the shape of electronic dispersion in the 2D Hubbard model. However, due to exact account for intracluster correlations, we are beyond the static approximation, which leads to manifestations of damping of excitations. We should note that in the Hubbard model, where an electron-phonon interaction is absent, we find the intracluster charge correlation functions to demonstrate a sharp decrease with distance without any special features, as shown in Appendix \ref{sec:a3}.

First, we have shown that in our CPT calculations the evolution of the Fermi surface from small to high doping proceeds through three stages, as within the generalized mean-field approximation\cite{Korshunov, Ovchinnikov09, Ovchinnikov11} and the DCA studies \cite{Gull09, Gull10}. At low doping SPG is observed accompanied by short-range AFM. As doping increases we observe a transition to WPG and destruction of short-range AFM. With further doping the pseudogap closes and the large Fermi surface is observed, the spin correlations are very weak in this case and restricted mainly to the first neighbors, which is typical for a paramagnetic non correlated state.

Next, we obtained the temperature dependent $\mathbf{k}$- and $\omega$-resolved spectral function within CPT and thus compared the doping and temperature evolutions of the pseudogap with each other within the same framework. Thus, we are able to draw the main conclusion: The electronic structure with temperature goes through the same stages as with doping due to a very similar behavior of spin correlations. This result is schematically depicted in Fig.~\ref{fig:13}(d) where we used two lines $T^*$ and $T^{*1}$. We note that conclusions about at least two critical temperatures above $T_c$ follow from the analysis of recent experimental data \cite{Kordyuk, Vishik18}.

\begin{acknowledgments}
The reported study was funded by RFBR according to the research project No. 18-32-00256 (all the results concerning the influence of short-range correlations on the electronic spectral properties). The reported study was also funded by RFBR according
to the research project No. 18-32-01062; Russian Foundation for Basic Research and Government of Krasnoyarsk Territory, Krasnoyarsk Regional Fund of Science to the research projects ``Electronic correlation effects and multiorbital physics in iron-based materials and cuprates'' number 19-42-240007, and ``Features of electron-phonon coupling in high-temperature superconductors with strong electron correlations'' number 18-42-240017.
\end{acknowledgments}

\appendix
\section{\label{sec:a1} Details of CPT implementation}
\begin{figure}
\includegraphics{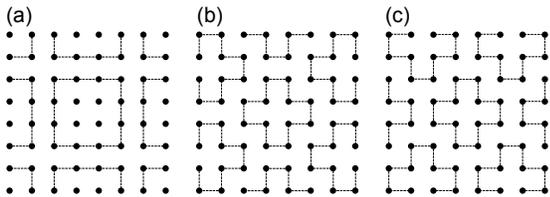}
\caption{\label{fig:ap1} Cluster coverings used for (a) a $4\times4$ cluster, (b),(c) a 12-site cluster. In the case of a 12-site cluster the hopping matrix was averaged among two coverings (b) and (c) similar to calculations of Ref.~\onlinecite{Nikolaev12}.}
\end{figure}

To implement CPT in this paper we follow the general logic of the X-operator perturbation theory \cite{Zaitsev75, Izyumov91, book_Valkov}. The lattice is covered by translations of a cluster (see Fig.~\ref{fig:ap1}), and Eq.~\ref{eq:1} is rewritten as
\begin {equation}
H = H_c + H_{cc},
\label {eq:A1}
\end {equation}
where $H_c$ is the intracluster part and $H_{cc}$ is the intercluster part
\begin {equation}
H_{cc} = \sum_{\mathbf{f},\mathbf{r},i,j}{T^{\mathbf{r}}_{i,j}a^{\dag}_{\mathbf{f},i}a^{}_{\mathbf{f}+\mathbf{r},j}},
\label {eq:A2}
\end {equation}
where $\mathbf{f}$ runs over clusters, $\mathbf{r}$ labels neighbor clusters, $i$ and $j$ are indices of sites within a cluster, and spin index is omitted here and below. We define the Green's functions 
\begin {equation}
D_{\alpha,\beta}\left(\mathbf{\tilde{k}},\omega\right) = \allowbreak {\left\langle \left\langle X^{\alpha}|{X^{\beta}}^{\dag} \right\rangle \right\rangle}_{\mathbf{\tilde{k}},\omega},
\label {eq:A3}
\end {equation}
where $\mathbf{\tilde{k}}$ is the wave vector defined in the reduced Brillouin zone. The fermionic Hubbard operators in Eq.~\ref{eq:A3},
\begin {equation}
X^{\alpha} = X^{p,q} = \left|p\right\rangle\left\langle q\right|,
\label {eq:A4}
\end {equation}
are supposed to be built on the full basis of cluster eigenstates denoted as $\left|p\right\rangle$ and $\left|q\right\rangle$ so that if $\left|p\right\rangle$ is a state with $n-1$ particles then $\left|q\right\rangle$ is a state with $n$ particles. Using the fact that an annihilation operator of an electron on a site $i$ belonging to a cluster $\mathbf{f}$ is a linear combination of $X$-operators,
\begin {equation} 
a_{\mathbf{f},i}=\sum_{\alpha}{\gamma_{i}\left(\alpha\right)X^{\alpha}_{\mathbf{f}}},
\label {eq:A5}
\end {equation} 
where $\gamma_{i}\left(\alpha\right)$ are the annihilation operator's matrix elements, the Hamiltonian given by Eq.~\ref{eq:A1} can now be written in terms of the Hubbard operators:
\begin {equation}
H= \sum\limits_{\mathbf{f},m}{E_m} X^{mm}_{\mathbf{f}} + \sum\limits_{\mathbf{f},\mathbf{r}}\sum\limits_{\alpha,\beta}{V^{\mathbf{r}}_{\alpha,\beta}{X_{\mathbf{f}}^{\alpha}}^{\dag}X_{\mathbf{f} + \mathbf{r}}^{\beta}},
\label {eq:A6}
\end {equation}
where $m$ runs over all cluster eigenstates, the intercluster coefficients are
\begin {equation}
V^{\mathbf{r}}_{\alpha,\beta} = \sum_{i,j}{\gamma^{*}_{i}\left(\alpha\right)\gamma^{}_{j}\left(\beta\right)T^{\mathbf{r}}_{i,j}}.
\label {eq:A7}
\end {equation} 
The Dyson equation in terms of the Hubbard operators reads \cite{book_Valkov}:
\begin {eqnarray}
\hat{D}\left(\mathbf{\tilde{k}},\omega\right) 
= [{\hat{D^0}\left(\omega\right)}^{-1}
&-&\hat{P}\left(\mathbf{\tilde{k}},\omega\right)\hat{V}\left(\mathbf{\tilde{k}}\right) \nonumber \\
&+& \hat{\Sigma}\left(\mathbf{\tilde{k}},\omega\right)
 ]^{-1}\hat{P}\left(\mathbf{\tilde{k}},\omega\right),
\label {eq:A8}
\end {eqnarray} 
where all the matrices are defined in terms of band indices $\alpha$ and $\beta$,
\begin {equation}
V_{\alpha \beta}\left(\mathbf{\tilde{k}}\right) = \sum\limits_{\mathbf{r}}
V^{\mathbf{r}}_{\alpha\beta}e^{i\mathbf{\tilde{k}}\mathbf{r}}
\label {eq:A9}
\end {equation}
is the element of the hopping matrix and  
\begin {equation}
{D^0_{\alpha,\beta}\left(\omega\right)} = \frac{\delta_{\alpha,\beta}}{\omega - E_{\alpha} + \mu}
\label {eq:A10}
\end {equation}
is the exact local propagator, $E_{\alpha} = E_{q} - E_{p}$, and $\mu$ is the chemical potential. In Eq.~\ref{eq:A8}, $\hat{\Sigma}\left(\mathbf{q},\omega\right)$ is the intercluster self-energy and $\hat{P}\left(\mathbf{\tilde{k}},\omega\right)$ is the strength operator. In the Hubbard-I approximation for the intercluster hopping one has $\hat{\Sigma}\left(\mathbf{\tilde{k}},\omega\right) = 0$ and $P_{\alpha \beta}\left(\mathbf{\tilde{k}},\omega\right) = \delta_{\alpha \beta} F_{\alpha}$, where
\begin {equation}
F_{\alpha} =  \left\langle X^{pp} \right\rangle + \left\langle X^{qq} \right\rangle = n_p + n_q,
\label {eq:A11}
\end {equation}
where the diagonal averages $\left\langle X^{pp}\right\rangle$ and $\left\langle X^{qq}\right\rangle$ are the occupancies $n_p$ and $n_q$ of cluster energy levels. Thus, the electronic structure in this approximation is defined by the equation
\begin {equation}
\hat{D}\left(\mathbf{\tilde{k}},\omega\right)^{-1} = \left[\hat{F}\hat{D^0}\left(\omega\right)\right]^{-1} - \hat{V}\left(\mathbf{\tilde{k}} \right).
\label {eq:A12}
\end {equation}

The diagonal matrix $\hat{F}$ in Eq.~\ref{eq:A12}, which consists of all levels' occupancies, ideally should be calculated via Eq.~\ref{eq:A3}, but, since it is an extremely cumbersome task, we use the following approximation to work with a fixed particle number. We set a number of electrons per cluster $n_e$ by assuming non zero occupations $1-x$ and $x$ for only two sectors of the Hilbert space with $n$ and $n-1$ electrons, respectively, so that 
\begin {equation}
n_e = \left(1-x\right)n + x (n - 1),
\label {eq:A13}
\end {equation}
where $n$ is the integer number of electrons allowed by a finite cluster such that $n-1 < n_e < n$. For example, doping $p=0.0625=1/16$ or $p = 1/6$ for a $4 \times 4$ cluster is obtained by choosing $n = 15$ and $x = 0$ or $n = 14$ and $x = 2/3$, respectively. Then we calculate the occupation numbers $n_p$ for the sector with $n-1$ electrons and $n_q$ for the sector with $n$ electrons within a canonical ensemble for each of them:
\begin {eqnarray}
n_p & = & \frac{x}{Z_{n-1}}\exp{\left(-\beta E_p\right)}, \nonumber \\
n_q & = & \frac{1-x}{Z_{n}}\exp{\left(-\beta E_q\right)},
\label {eq:A14}
\end {eqnarray}
where $Z_{n}$ is a canonical partition function for a cluster with $n$ electrons.

For practical calculations with relatively large clusters used in this study, using Eq.~\ref{eq:A5} one can obtain from Eq.~\ref{eq:A12} an analogous equation written in terms of matrices defined in cluster sites' indices  as in Ref.~\cite{Senechal00}:
\begin {equation}
\hat{G}\left(\mathbf{\tilde{k}},\omega\right)^{-1} = \hat{G}^c\left(\omega\right)^{-1} -\hat{T}\left(\mathbf{\tilde{k}}\right),
\label {eq:A15}
\end {equation}
where 
\begin {equation}
G^c_{i,j}\left(\omega\right)=\sum_{\alpha, \beta}{\gamma_{i}\left(\alpha\right)\gamma^*_{j}\left(\beta\right)F_{\alpha}D^0_{\alpha,\beta}\left(\omega\right)}
\label {eq:A16}
\end {equation}
and $\hat{T}\left(\mathbf{\tilde{k}}\right) = \sum\limits_{\mathbf{r}}\hat{T}^{\mathbf{r}}e^{i\mathbf{\mathbf{\tilde{k}}}\mathbf{r}}$. 

There exists a number of methods designed to calculate efficiently finite-temperature properties of a cluster. For example, finite temperature Lanczos extensions \cite{Jaklic94, Aichhorn03, Long03, Okamoto18} or thermal pure quantum state (TPQ) methods \cite{Okamoto18, Sugiura12, Steinigeweg14}. In this work the matrix elements defined in Eq.~\ref{eq:A5}, which enter Eq.~\ref{eq:A16}, were calculated explicitly from the (quite large) set of the lowest-lying eigenstates (typically, 6400 for each subsector of the Hilbert space with a given particle number and spin projection) obtained using a numerically robust Lanczos method modification \cite{Wu00} realized in the Scalable Library for Eigenvalue Problem Computations (SLEPc) \cite{Hernandez05}. This approach is quite expensive numerically, but affordable in CPT, where no iterative diagonalization of the cluster Hamiltonian is needed, and possesses no statistical or systematical errors apart from controllable neglecting the highest-energy excitations with minor spectral weight. At zero temperature Eq.~\ref{eq:A16} is equivalent to a linear combination of cluster Green functions calculated using Lanczos method starting from the ground state eigenvector of each subsector multiplied by the corresponding occupation number. Each of the eigenvectors of the ground state served as a starting vector in the Lanczos procedure to contribute to the cluster Green function. A finite Lorentzian broadening $\delta=0.16t$ was used to calculate the Green function.

\section{\label{sec:a2} Comparison of results with 16-site and 12-site clusters}
In the main section we discussed results obtained at $T=0$ with a $4\times4$ cluster. When discussing finite-temperature results we have to restrict ourselves by a 12-site cluster. In this section we compare the results on electronic spectrum and spin-spin correlation functions obtained with 16-site and 12-site clusters at $T=0$, $p = 0.167$.

\begin{figure}
\includegraphics{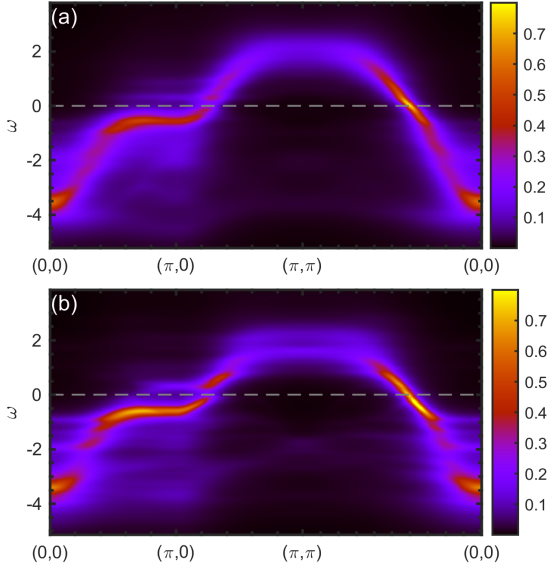}
\caption{\label{fig:ap2} The electronic spectral function obtained at $T=0$ and $p=0.167$ with (a) a $4\times 4$ cluster and (b) a 12-site cluster.}
\end{figure}

\begin{figure}
\includegraphics{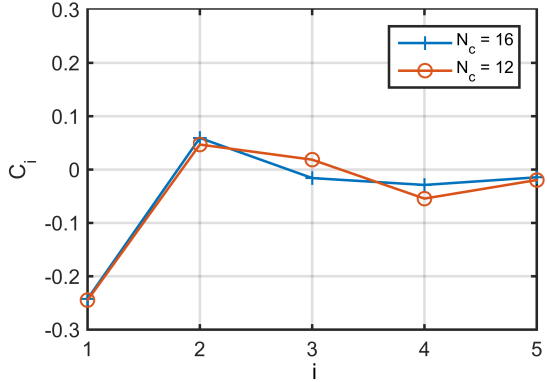}
\caption{\label{fig:ap3} Spin-spin correlation functions (Eq.~\ref{eq:2}) obtained with two types of cluster at $T=0$ and $p=0.167$.}
\end{figure}

From Fig.~\ref{fig:ap2} a qualitative agreement in dispersion shape between the spectra obtained with 12-site and 16-site clusters is seen. However, while the most general features agree, on the finer scale there are some disagreements. The pseudogap is clearly more pronounced in the case of a 12-site cluster. The analysis of spin correlations in Fig.~\ref{fig:ap3} shows that for the chosen value of doping the short-range AFM is still present at least till the fourth coordination sphere within a 12-site cluster, but it is violated at the third and further within a 16-site one. For the first two coordination spheres the quantitative agreement is observed. We note that for $p=0.167$ the values of $C_i$ on a 16-site cluster were estimated in the same manner as we did for the cluster Green function in CPT by choosing the weight factors. Qualitatively, we conclude from spin correlations and spectral function that cluster shape and size effect leads to SPG for a 12-site cluster and a WPG for 16-site cluster at this doping and temperature.

\section{\label{sec:a3} Short-range charge correlations}

\begin{figure}
\includegraphics{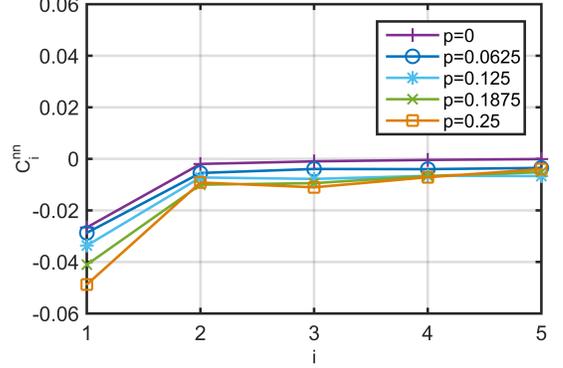}
\caption{\label{fig:ap4} Charge correlation functions (Eq.~\ref{eq:A17}) obtained within a $4\times 4$ cluster for different doping values as indicated in the inset for $U = 8$, $t'=-0.2$, and $t'' = 0$.}
\end{figure}

In the main part of the paper the dependance of the electronic structure on spin correlations was discussed, since in the absence of phonons there is no tendency to a CDW formation within a cluster has been observed. This is demonstrated in Fig.~\ref{fig:ap4}, where the charge correlations
\begin {equation}
C^{nn}_i = \left\langle \left(n_a-\left\langle n_a \right\rangle\right)\left(n_b-\left\langle n_b \right\rangle\right)\right\rangle_{\left|\mathbf{R}_a-\mathbf{R}_b\right|\in i},
\label {eq:A17}
\end {equation}
are shown in analogy with spin correlations. These charge correlations have the same sign for all coordination spheres and only for the first sphere their value is significantly different from zero.

\bibliography{paper}

\end{document}